\documentclass[11pt,twoside]{article}

\usepackage{asp2004}
\usepackage{epsf}
\usepackage{psfig}
\usepackage{lscape}

\begin{document}
\title{Rotation in the ZAMS: Be and Bn stars}
\author{J. Zorec$^1$, Y. Fr\'emat$^2$, C. Martayan$^3$, L. Cidale$^4$, A.
Torres$^4$} 
\affil{$^1$Institut d'Astrophysique de Paris, UMR7095 CNRS, Univ. P\&M Curie}
\affil{$^2$Royal Observatory of Belgium}
\affil{$^3$Observatoire de Paris-Meudon, France}
\affil{$^4$ Facultad de Ciencias Astron\'omicas y Geof\'\i sicas, UNLP, 
Argentina}

\begin{abstract}
 We show that Be stars belong to a high velocity tail of a single B-type 
star rotational velocity distribution in the MS. This implies that: 1) the 
number fraction N(Be)/N(Be+B) is independent of the mass; 2) Bn stars having 
ZAMS rotational velocities higher than a given limit might become Be stars.
\end{abstract}

% \vspace{-0.5cm}
\vspace{-0.7cm}

\section{ZAMS rotational velocities of Be stars}\label{zrv}  

 The observed $V\!\sin i$ of 127 galactic Be stars were corrected for 
uncertainties due to the gravitational darkening effect with the {\sc 
fastrot} models (Fr\'emat et al. 2005). Using the same models we calculated 
the today true rotational velocity $V$ of each star by determining its 
inclination angle $i$. With evolutionary tracks for rotating stars we 
determined the mass and age of the studied stars (Zorec et al. 2005). 
From each $V$ we estimated the corresponding $V_{\rm ZAMS}$ by taking into 
account four first order effects affecting the evolution of equatorial 
velocities: 1) variation induced by the time-dependent stellar radius; 2) 
angular momentum (AM) loss due to mass-loss phenomena in stars with mass 
$M\ga15M_{\odot}$ and conservation of AM for $M\la15M_{\odot}$; 3) changes of
the inertial momentum due to evolutionary effects and to the rotation using 2D 
barotropic models of stellar structure (Zorec et al. 1988); 4) internal
redistribution of the AM in terms of the meridional circulation time scale 
parameterized from Meynet \& Maeder's (2000) models. The obtained $V_{\rm 
ZAMS}$ against the stellar mass are shown in Fig.~{\ref{f1}a, where there is a 
neat mass-dependent limiting cut $V_{\rm min}(M)$ that confines all studied Be 
stars in a high velocity sector. This indicates that stars need to have 
$V_{\rm ZAMS}\ga$ $V_{\rm min}$ to become Be in the MS phase. For each 
mass-interval we divided the $V_{\rm ZAMS}$ by $V_{\rm min}$ and obtained the
global histogram shown in Fig.~{\ref{f1}b. The fit that better describes the 
distribution of $V_{\rm ZAMS}(M)/V_{\rm min}(M)$ obtained is a Gaussian tail.
The histogram concerns only 17\% roughly of stars out of the whole B star MS 
population. Since more than 80\% of MS B-type stars must then be in the 
$V_{\rm ZAMS}/V_{\rm min}$ $\la 1$ interval, it implies that Be stars do not 
form a separate distribution, but possibly a tail of a general distribution 
that encompasses the whole B-star MS population.

\begin{figure}[t]
\centerline{\psfig{file=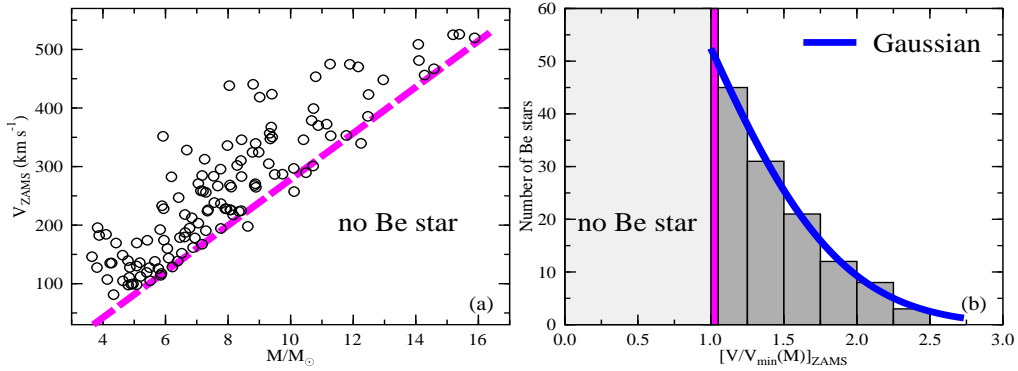,width=13.5truecm,height=5truecm}}
% \centerline{\psfig{file=jzorec1_fig1.eps,width=12.5truecm,height=4truecm}}
\caption[]{a) Distribution of the equatorial velocities in the ZAMS of the 
studied Be stars against the mass . b) Frequency distribution of $V_{\rm 
ZAMS}/V_{\rm min}$ and fit with a Gaussian function. No Be star has $V_{\rm 
AMS}/V_{\rm min}<1$}
\label{f2}
\end{figure}

\begin{figure}[h!]
\centerline{\psfig{file=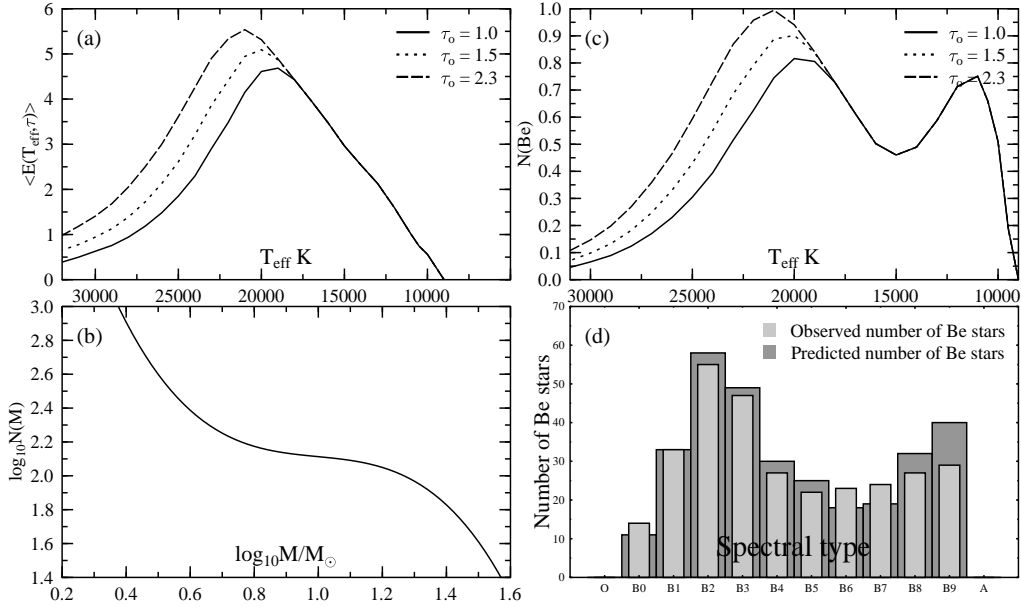,width=13.5truecm,height=8truecm}}
% \centerline{\psfig{file=surf359_j.eps,width=7truecm,height=9.2truecm}}
\caption[]{a) H$\alpha$ emission intensity function as a function of $T_{\rm 
eff}$ and $\tau_o$. b) Number of stars around the Sun in an apparent V=7
magnitude limited volume. c) Normalized predicted number of Be stars near the 
Sun. d) Comparison of predicted with observed number of Be stars near the Sun}
\label{f1}
\end{figure}

\vspace{-0.5cm}

\section{Mass-independent frequency of Be stars }\label{mif}  

 The number of detected Be stars is given by: ${\rm d}N(Be,i)\propto P_{\rm 
Be}E_{H\alpha}(T_{\rm eff},\tau,i)\phi(\tau)$ $\times N(T_{\rm eff})\sin i{\rm
d}\tau{\rm d}i$, where $P_{\rm Be}$ is the mass-independent probability of 
becoming Be star (see Sect.~\ref{zrv}); $E_{H\alpha}$ is the probability of 
detecting a Be star measured in terms of the intensity of the H$\alpha$ 
emission produced by a disc with opacity $\tau$ shown in Fig.~\ref{f2}a; 
$N(T_{\rm eff)}$ is the total number of stars with a given $T_{\rm eff}$ given
by the IMF function; $\sin i$ is the probability of seeing the disc at 
inclination $i$. We can integrate the indicated relation for an apparent V=7 
magnitude limited volume to represent the fraction of Be stars near the Sun. 
The reduced apparent magnitude IMF function is shown in Fig.~\ref{f2}b. It 
comes then that the product of $E_{H\alpha}$ with the `local' IMF produces the 
curves shown in Fig.~\ref{f2}c, which represent the normalized predicted 
number $N(Be)$ of Be stars. The comparison of the predicted ($\tau_o\!=\!1$) 
with the actually observed number of Be stars around the Sun is shown in 
Fig.~\ref{f2}d. This shows that the apparently bi-modal distribution of Be 
stars around the Sun against the spectral type is entirely explained by the 
interplay of the probability of detecting Be stars ($E_{H\alpha}$), the shape
of the IMF and $P_{\rm Be}(M)=$ constant. 

\section{Some Bn stars might be progenitors of Be stars}\label{bn}

 Figure~\ref{f3}a shows the apparent V=7 magnitude limited counts of dwarf Be
stars relative to dwarf B stars. There is an apparent lack of dwarf Be stars 
cooler than spectral type B7. This could be due to genuine Be stars whose 
discs are minute and/or too cool for the H$\alpha$ emission be detectable
and/or, to fast rotating B stars that still had not attained the required 
properties to become fully-fledged Be stars. According to findings in
Sect.~\ref{zrv}, which are supported by those in Sect.~\ref{mif}, the relation
shown in Fig.~\ref{f3}a should be straightened, so that $\ln{N(Be)}/\ln{N(B)}=$
constant. It has been shown in Zorec et al. (2005) that there is a lack of Be
stars with masses $M \la 7M_{\odot}$ in the first half of the MS phase. 
However, Fig.~\ref{f3}b shows that apparent V=7 magnitude limited counts of Bn
stars increase strongly at spectral types cooler than B7. Since most of them 
are dwarfs, they probably had not enough time to attain the angular velocity
ratio $\Omega/\Omega_c\simeq0.9$ that characterizes Be stars (Fr\'emat et al.
2005). The determination of the $V_{\rm ZAMS}$ of 100 Bn stars is underway. 
This will enable us to see which of them has the required condition $V_{\rm 
ZAMS}\ga V_{\rm lim}$ to become Be. Other related subjects can be found in 
http://www2.iap.fr/users/zorec/. 

\begin{figure}[]
\centerline{\psfig{file=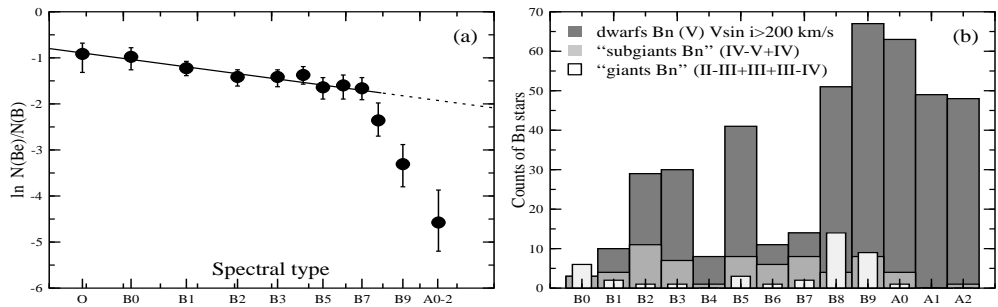,width=13.5truecm,height=4truecm}}
\caption[]{a) Apparent magnitude limited counts (V=7mag) of dwarf Be stars 
relative to dwarf B stars. b) Apparent magnitude limited counts (V=7mag) of 
Bn stars. Most Bn stars are dwarfs, which may indicate that they are fast 
rotators in the first MS evolutionary phases}
\label{f3}
\end{figure}

%\\subsection{}   
%\subsubsection{}   %%% Lowest level section head (remove "%" symbol)
%\section*{}	%%% Unnumbered top level section head (remove "%" symbol)
%\subsection*{}   %%% Unnumbered second level section head (remove "%" symbol)

% \acknowledgements %%% Text of acknowledgements runs on after this command.

\end{document}